# Detection of the Vortex Dynamic Regimes in MgB$_2$ by Third Harmonic AC Susceptibility Measurements

C. Senatore[1,2], M. Polichetti[1], N. Clayton[2], R. Flükiger[2], S. Pace[1]

[1]Dipartimento di Fisica "E.R. Caianiello" and INFM, Università di Salerno, Italy
[2]Département de Physique de la Matière Condensée, Université de Genève, Switzerland

**Abstract**

In a type-II superconductor the generation of higher harmonics in the magnetic response to an alternating magnetic field is a consequence of the non-linearity in the *I-V* relationship. The shape of the current-voltage (*I-V*) curve is determined by the current dependence of the thermal activation energy $U(J)$ and is thus related to the dynamical regimes governing the vortex motion. In order to investigate the vortex dynamics in MgB$_2$ bulk superconductors we have studied the fundamental ($\chi_1$) and third ($\chi_3$) harmonics of the ac magnetic susceptibility. Measurements have been performed as a function of the temperature and the dc magnetic field, up to 9 T, for various frequencies and amplitudes of the ac field. We show that the analysis of the behaviour in frequency of $\chi_3(T)$ and $\chi_3(B)$ curves can provide clear information about the non-linearity in different regions of the *I-V* characteristic. By comparing the experimental curves with numerical simulations of the non-linear diffusion equation for the magnetic field we are able to resolve the crossover between a dissipative regime governed by flux creep and one dominated by flux flow phenomena.

# 1. Introduction

The discovery of the new superconductor MgB$_2$ with $T_c$ = 39 K has stimulated a lot of interest in both fundamental and applied research [1]. The physical properties of this binary compound are intermediate between those of conventional and high-$T_c$ superconductors. Due to its large coherence length ($\xi_0$ = 50 Å), MgB$_2$ has weak-link-free grain boundaries [2] and a high transport critical current density $J_c$ [3]. However, the rapid decrease of $J_c$ with applied magnetic field limits the potential of MgB$_2$ for high-field power applications. Studying the pinning properties and vortex dynamics therefore represents a basic means to understand the mechanism governing magnetic irreversibility. One of the most popular methods to investigate vortex dynamics is the measurement of the response of the vortex lattice to an ac magnetic field [4]. Ac susceptibility is a powerful tool to study the different dynamical regimes of the flux lines; this technique allows us to induce changes in the vortex dynamics by changing the external parameters, such as the ac field frequency and amplitude, the dc magnetic field intensity and the temperature.

The response of type-II superconductors to an ac magnetic field can be either linear or non-linear. The linear response can be divided into two different regimes: the flux flow (FF) regime and the thermally activated flux flow (TAFF) regime. In the FF regime, the response is dominated by a viscous motion of the vortex liquid [5]. In the TAFF regime, thermally activated vortex jumps between favourable metastable states of the vortex lattice come into play [6] and contribute to the ac response, especially near $T_c$. In both of these cases the linear response is due to an Ohmic resistive state $E = \rho J$ in the sample, where $\rho$ ($B,T$) is independent of $J$. Therefore, in the presence of a dc field greater than the ac field, the electrodynamics of a superconductor in these regimes can be described in terms of a normal metal, governed by the skin effect. However, the magnetic field dependence of the two resistivities $\rho_{FF}$ and $\rho_{TAFF}$ can induce, in the presence of only an ac field and not a dc field, a non-linear behaviour and thus higher harmonics in the ac response.

When the vortex motion is governed by the thermally activated flux creep (FC) phenomenon, the resistivity of the sample is expressed by the $J$-dependent thermally activated form:

$$\rho(J) = \rho_0 \, exp\left[-\frac{U(J)}{k_B T}\right] \qquad (1)$$

with $U(J)$ being the effective activation energy. Therefore, the *E-J* characteristic exhibits non-linear behaviour in the flux creep regime, which is a direct consequence of the current dependence of $U$; this leads to the generation of higher harmonics in the ac susceptibility.

The third harmonic response of the ac susceptibility is highly sensitive to the different flux-dynamic regimes. Nevertheless, the interpretation of the experimental results requires the use of theoretical models, which relate the ac magnetic response to the electrical transport properties. The first interpretation comes from the original Bean critical state model [7,8], which attributes the harmonic generation to the hysteretic relationship between the magnetization and the external field due to flux pinning. Since the harmonics of the ac susceptibility calculated using the critical state approach are independent of the frequency of the applied ac field [9], the simple critical state description is unable to explain the frequency dependence clearly observed in many experiments [10]. In order to account for this behaviour, some authors have integrated analytically or numerically the non-linear diffusion equation for the magnetic field [11-14], taking into account the different dissipative phenomena (FC, FF, TAFF) responsible for the diffusion processes.

In this article we investigate the dynamical regimes governing the vortex motion in $MgB_2$ bulk samples by measuring the fundamental ($\chi_1 = \chi_1' + i\chi_1''$) and third ($\chi_3 = \chi_3' + i\chi_3''$) harmonics of the ac magnetic susceptibility as a function of the temperature and the dc magnetic field. In particular we have determined the connection between the different structures that appear in the $\chi_3(T)$ curves with increasing frequency and the dissipative phenomena responsible for the magnetic response of the sample. This has been obtained by comparing the experimental curves with numerical simulations of the non-linear magnetic diffusion equation. A new empirical model for the magnetic diffusion process has been introduced in order to describe the crossover from the flux creep to the flux flow regime.

This paper is organized as follows. In Sec.2 we describe the numerical method applied to resolve the non-linear magnetic diffusion problem. The experimental procedure and the $MgB_2$ sample are described in Sec.3. In Sec.4 the $\chi_3(T)$ and $\chi_3(B)$ measurements are discussed and compared with the calculated curves. Finally, Sec.5 is dedicated to conclusions.

## 2. The non-linear diffusion equation

The diffusion equation which governs the penetration of the magnetic flux in a type-II superconductor can be obtained as follows. Since the electric field induced by the flux motion is related to the current density by Ohm's law $E = \rho(B,J,T) J$, by using the Maxwell equations $\nabla \times E = -\partial B /\partial t$ and $\nabla \times B = \mu J$ (neglecting the displacement currents), one obtains the equation describing the flux diffusion:

$$\frac{\partial \boldsymbol{B}}{\partial t} = -\nabla \times \left[ \frac{\rho(B,J,T)}{\mu} \nabla \times \boldsymbol{B} \right]. \qquad (2)$$

In the case of an infinite slab of thickness $d$ in a parallel field geometry Eq.(2) can be rewritten as:

$$\frac{\partial B}{\partial t} = \frac{\partial}{\partial x}\left[\frac{\rho(B,J,T)}{\mu}\frac{\partial B}{\partial x}\right] \qquad (3)$$

with the boundary conditions $B(\pm d/2,t) = B_{ext}(t) = B_{dc} + B_{ac} \sin(2\pi f t)$. The diffusion coefficient $\rho(B,J,T)$ is the residual resistivity due to the flux motion and is related to the vortex dynamic regimes. In order to account for changes in the non-linear behaviour produced by variations of the currents induced by the ac driving field, we describe the diffusivity in terms of a new empirical model incorporating both flux creep and flux flow resistivities.

Measurements of the $I$-$V$ characteristics of traditional low-$T_c$ superconductors [15,16] show that no detectable voltage appears until $I$ exceeds the critical current. If the current is increased beyond the depinning threshold the observed voltage is linear in current and the slope $\Delta V /\Delta I$ is interpreted as the flux flow resistance. This means that the differential resistivity $d\rho (\propto dV/dI)$ is simply a step-function (Fig. 1):

$$d\rho = \frac{dE}{dJ} = \rho_{FF}\,\Theta(J - J_c) = \rho_n \frac{B}{B_{c2}}\Theta(J - J_c). \qquad (4)$$

Eq.(4) is only true for low-$T_c$ superconductors in which the thermally activated flux creep phenomenon is negligible. In order to take into account the effects of thermal activation in the vortex motion we propose the following empirical expression for the differential resistivity (Fig. 1):

$$d\rho = \frac{dE}{dJ} = \rho_{FF} \frac{1}{\frac{k_B T}{U_P(B,T)} \exp\left[\frac{U_P(B,T)}{k_B T}\left(1 - \frac{J}{J_c(B,T)}\right)\right] + 1}, \quad (5)$$

where $U_P(B,T)$ is the activation energy, $J_c(B,T)$ the critical current density, $k_B$ the Boltzmann constant and $T$ the temperature. In the limit $T \to 0$, Eq.(5) reduces to the step-function behaviour of Eq.(4):

$$\text{for } J < J_c \quad \exp\left[\frac{U_P}{k_B T}\left(1 - \frac{J}{J_c}\right)\right] \xrightarrow{T \to 0} \infty \Rightarrow \frac{dE}{dJ} \xrightarrow{T \to 0} 0 \quad (6a)$$

$$\text{for } J > J_c \quad \exp\left[\frac{U_P}{k_B T}\left(1 - \frac{J}{J_c}\right)\right] \xrightarrow{T \to 0} 0 \Rightarrow \frac{dE}{dJ} \xrightarrow{T \to 0} \rho_{FF} \quad (6b)$$

In the limit $J \to 0$ the flux creep behaviour [14] is recovered:

$$\lim_{J \to 0} \frac{dE}{dJ} = \rho_{FF} \frac{U_P}{k_B T} \exp\left[-\frac{U_P}{k_B T}\left(1 - \frac{J}{J_c}\right)\right] = \frac{dE_{creep}}{dJ}, \quad (7)$$

while in the limit $J \to \infty$ the flux flow resistivity is recovered.

By integrating Eq.(5) with the initial condition $E(J=0) = 0$, we obtain the analytical form of the $E(J)$ curve:

$$E(J) = \rho_{FF} \frac{k_B T}{U_P(B,T)} J_c(B,T) \ln\left[\frac{1 + \frac{U_P(B,T)}{k_B T} \exp\left[-\frac{U_P(B,T)}{k_B T}\left(1 - \frac{J}{J_c(B,T)}\right)\right]}{1 + \frac{U_P(B,T)}{k_B T} \exp\left(-\frac{U_P(B,T)}{k_B T}\right)}\right]. \quad (8)$$

The diffusion coefficient which results from Eq.(8) has been used to perform the numerical simulations reported in this paper. Within this approach different regimes of flux motion are smoothly connected in the $E(J)$ characteristic and the crossover between FC and FF occurs as the frequency of the ac driving field is increased.

The non-linear diffusion equation for the magnetic field has been numerically solved by means of Fortran NAG [17] routines. The algorithm computes the time evolution of the local field profile by integrating a discrete version of Eq.(3) using Gear's method for a fixed number of spatial meshes. In order to obtain the harmonic susceptibilities $\chi_n = \chi_n' + i\chi_n''$, we have first calculated the magnetization $M$ for the applied time-dependent field, i.e., the magnetization loop, and then its Fourier transforms.

In order to account for the temperature and field dependence of the ac susceptibilities we have to specify the temperature and field dependence of the critical current density $J_c(B,T)$ and the thermal activation energy $U_P(B,T)$. To this end, we chose for the temperature dependences the following forms:

$$J_c(B=0,T) = J_c(B=0,T=0)\left(1-\tau^2\right)^{\frac{5}{2}}\left(1+\tau^2\right)^{-\frac{1}{2}} \tag{9a}$$

$$U_P(B=0,T) = U_P(B=0,T=0)\left(1-\tau^4\right), \tag{9b}$$

which are the predictions of the collective pinning (CP) model [6,18]. In Eqs.(9a) and (9b) $\tau$ is the reduced temperature $T/T_c$ and $T_c$ is the critical temperature of the sample.

The field dependences of $J_c$ and $U_P$ have been determined by means of DC magnetization (Vibrating Sample Magnetometer) measurements performed on the same $MgB_2$ sample used to study the harmonic response. The resulting dependences are:

$$J_c(B,T=0) = J_c(B=0,T=0)\exp(-\alpha B) \tag{10a}$$

$$U_P(B,T=0) = U_P(B=0,T=0)\exp(-\beta B), \tag{10b}$$

with $\alpha = 1.4$ and $\beta = 0.5$. The upper critical field is [19]:

$$B_{c2}(T) = B_{c2}(T=0)\frac{1-\tau^2}{1+\tau^2}, \tag{11}$$

with $B_{c2}(T=0) = 18$ T. The other parameters used for all the simulations reported in this paper are $U_P(B=0,T=0)/k_B T_c = 100$, $J_c(B=0,T=0)\times d = 0.75\times 10^6$ A/m, $\rho_n = 2\times 10^{-6}$ $\Omega$cm.

Equations and results have been expressed in SI units. We recall that in SI units volume susceptibilities are dimensionless.

## 3. Experimental details

The ac susceptibility measurements reported in this work have been performed on a polycrystalline $MgB_2$ bulk sample prepared by reactive liquid infiltration [20]. The temperature dependences of both the first and the third harmonics have been measured by sweeping the temperature at a rate of $\Delta T/\Delta t$ = 0.4 K/min from 4.2 to 45 K. We have also measured the magnetic field dependence of $\chi_1(B)$ and $\chi_3(B)$ sweeping the dc field from 0 to 9 T, at different temperatures (20, 25, 30 K). Measurements have been performed using different ac field amplitudes (1, 2, 4, 8, 16 G) in the frequency range 7 ÷ 5007 Hz. Both the ac and dc fields are applied parallel to the longest side of the sample, whose dimensions are 1.59×0.34×15 mm$^3$. The sample exhibits a sharp transition with a $T_c$ = 38.8 K, evaluated from the real part of $\chi_1(T)$ measured for $B_{ac}$ = 1 G and $f$ = 5007 Hz; the transition width, estimated by the 10%-90% criterion, is about 0.7 K.

Because of the sensitivity of the third harmonic response to the phase setting of the measurements, we have set the phase in a very accurate way. We have measured the wide-band susceptibility [4,21] as a function of the lock-in phase in the range [-180°, 180°] for each measured frequency at $B_{ac}$ = 1 G and $T$ = 4.2 K, both with and without the sample. We then calculate the difference between the curves acquired with and without the sample in order to remove all the contributions due to spurious signals. The amplitude susceptibility $\chi_a$ is proportional to the sample magnetization when the external field reaches its maximum value, whereas the remanent susceptibility $\chi_r$ is proportional to the magnetization remaining in the sample at the zero instantaneous value of the ac field. When the sample is in the Meissner state, $\chi_r$ is zero and $\chi_a$ reaches its maximum negative value. This allows us to determine the phase to an accuracy better than 0.1 degree.

## 4. Experimental results and discussion

### 4.1. *Third harmonics vs temperature*

Fig. 2 shows the temperature dependence of the real and imaginary components of the third harmonic $\chi_3$ measured with $f=1007$ Hz and $B_{ac}=4$ G. At temperatures higher than the peak temperature $T_P$ in $\chi_1''$, which corresponds in the critical state picture to the temperature at which the ac field is equal to the full penetration field, we can observe in the $\chi_3'(T)$ and $\chi_3''(T)$ curves the typical structures predicted by the Bean model, i.e., a positive peak that goes monotonously to zero in $\chi_3'(T)$ and a positive peak followed by a trough in $\chi_3''(T)$. At temperatures lower than $T_P$ ($\chi_1''$) the form of the curves deviates from the predictions of the critical state: a deep trough in $\chi_3'(T)$ and a bump before the main peak in $\chi_3''(T)$ indicate the presence of flux dynamic phenomena. In Fig. 3 we show the $\chi_3$ curves measured with $f=5007$ Hz and $B_{ac}=4$ G. By increasing the frequency, the shapes of $\chi_3'(T)$ and $\chi_3''(T)$ are substantially changed. The $\chi_3'$ peak exhibited at $f=1007$ Hz for temperatures higher than $T_P$ ($\chi_1''$) disappears and is replaced by the trough that is present just for $T<T_P$ ($\chi_1''$) in Fig. 2. In the $\chi_3''(T)$ curve, for $T>T_P$ ($\chi_1''$) the peak goes monotonously to zero, while a trough is formed from the bump found for $T<T_P$ ($\chi_1''$) in Fig. 2. The appearance of this new structure in $\chi_3''(T)$ is the main feature that emerges from our study of the frequency behaviour of the third harmonic response.

In Fig. 4 we report the $\chi_3''(T)$ experimental curves for $B_{ac}=4$ G and $f=1007, 1607, 2507, 3507, 5007$ Hz. Except for the curve measured at $f=1007$ Hz we notice that the height of the peak decreases as the frequency increases. The small trough for $T>T_P$ ($\chi_1''$) rapidly disappears and there is no trace of it for $f>1607$ Hz. This suggests the tendency of the sample to abandon the critical state condition. On the other hand, the behaviour in frequency of the bump/trough that arises for $T<T_P$ ($\chi_1''$) is well reproduced by the simulated curves reported in Fig. 5.

Numerical simulations have been performed in the framework of the model described in Sec.2, replicating the experimental conditions. We notice in Fig. 5a the appearance of the bump at $f$ =1007 Hz, and then of the trough ($f$ >1007 Hz) whose minimum reaches negative values in Fig. 5b. The minimum of the trough is always placed at a temperature lower than $T_P$ ($\chi_1''$) and this is in accordance with the experimental behaviour.

By increasing the frequency of the ac magnetic field, the electrical field induced in the sample is increased, which drives the system through regions of the *E-J* characteristic with different non-linearity. The good agreement of the simulated curves with the experimental data for $T < T_P$ ($\chi_1''$) allows us to state that the trough structure appearing in the $\chi_3''(T)$ curves is evidence of the progressive crossover between a dissipative regime governed by flux creep and one dominated by flux flow.

For $T > T_P$ ($\chi_1''$) the behaviour of the simulated curves does not agree with the experimental data. In particular a new peak appears in the simulations close to $T_c$, whose height increases with frequency. The disagreement with the experiment is likely to be due to the expression used in Eq.(8) for the field dependence of $\rho_{FF}$. In fact, it has been shown in the literature [22] that the expression $\rho_{FF}/\rho_n = B/B_{c2}$, as calculated by Bardeen and Stephen [5], fails to describe the experimental behaviour when $B \to B_{c2}$. In particular, in order to account for the interaction between the flux lines, the resistivity $\rho_{FF}(B)$ needs a quadratic correction, which has not been considered in the numerical simulations reported in this paper.

### 4.2. *Third harmonics vs dc magnetic field*

It has been stated in the literature [9] that measurements of harmonic susceptibilities as a function of any experimental variable (temperature, dc magnetic field, ac magnetic field) can be reduced to universal curves which describe the harmonic susceptibilities versus a single parameter which is a function of the full penetration field and of the ac applied field. This is strictly true when the critical state description holds and no vortex motion is present in the sample. It is well known that a dc magnetic field superimposed on an ac field modifies the dynamics of the flux lines and thus the harmonic response. In particular, if $B_{dc} \gg B_{ac}$, the flux flow regime becomes linear, and thus does not contribute to the signal of higher harmonics. Therefore the combined analysis of $\chi_3(T)$ and $\chi_3(B)$ allows us to detect the flux regimes corresponding to the structures appearing in the experimental curves.

Fig. 6a and 6b show the $\chi_3'(B)$ and $\chi_3''(B)$ curves measured at fixed temperature and frequency ($T = 20$ K and $f = 1607$ Hz) for different ac field amplitudes ($B_{ac} = 2, 4, 8, 16$ G) and Fig. 7a and 7b show the $\chi_3'(B)$ and $\chi_3''(B)$ curves measured at fixed temperature and ac field amplitude ($T = 20$ K and $B_{ac} = 4$ G) for different frequencies ($f = 1007, 1607, 3507, 5007$ Hz). In the $\chi_3'(B)$ curves (Fig. 6a and Fig. 7a) a trough structure in the region where the critical state description predicts $\chi_3' = 0$ appears, followed by the Bean peak at higher magnetic fields. This trough can be attributed to flux creep, as it is the only dynamic regime still non-linear even in presence of a magnetic field $B_{dc} >> B_{ac}$. The $\chi_3''(B)$ curves (Fig. 6b and Fig. 7b) show a peak followed by a trough. By increasing the ac field frequency and amplitude, the bump/trough structure, exhibited in the temperature sweeps (Fig. 4) in the region before the main peak, does not appear when sweeping the dc magnetic field. This further confirms that the bump/trough for $T < T_P(\chi_1'')$ in the $\chi_3''(T)$ curves is related to a change of the vortex dynamics toward a flux flow governed regime induced by temperature.

The numerical simulations performed in the framework of the model reported in Sec.2 for $T/T_c = 0.5$ and $f = 1607$ Hz (Fig. 8) agree well with the experimental behaviour of the $\chi_3(B)$ curves measured at different $B_{ac}$. In particular the variations in position and height of the structures of the curves are correctly reproduced. However, as can be noted by comparing the measured curves with the simulated ones, the values of the height/depth of the peaks/troughs are slightly different and this could be due to the values estimated for the parameters in the simulations. Moreover, our numerical calculations performed for $T/T_c = 0.5$ and $B_{ac} = 4$ G (Fig. 9) show that the height of the peaks increases in both the $\chi_3'(B)$ and $\chi_3''(B)$ simulated curves, while the depth of the troughs decreases in $\chi_3'$ and increases in $\chi_3''$ with increasing frequency; in the experiment we observe the opposite behaviour.

As shown in Ref. [14] the frequency behaviour of the $\chi_3$ peaks/troughs is related to the current dependence of the thermal activation energy $U(J)$. The general form for $U(J)$ is

$$U(J) = \frac{U_0}{\mu}\left[\left(\frac{J_c}{J}\right)^{\mu} - 1\right]. \qquad (12)$$

The choice $\mu = -1$ corresponds to the usual Kim-Anderson model [23] and has been used for the simulations reported in this paper (see Eq.(8)). On the other hand, the flux creep phenomena in the case of collective pinning/vortex glass [24-25] can be recovered from Eq.(12) by imposing $0<\mu<1$. The frequency behaviour encountered in the experimental $\chi_3(B)$ curves (Fig. 8), i.e. the decrease of the height of the peaks in $\chi_3'$ and $\chi_3''$, the increase of the depth of the trough in $\chi_3'$, the decrease of the depth of the trough in $\chi_3''$ as the frequency increases, agrees well with the simulations performed in Ref. [14] in the framework of the collective pinning/vortex glass.

In Fig. 10 we report the $|\chi_3|(B)$ curves measured for $f = 1607$ Hz and $B_{ac} = 4$ G at different temperatures ($T = 20, 25, 30$ K). As the temperature increases the thermally activated phenomena are enhanced. In fact, we observe in the inset of Fig. 10 that the trough in the $\chi_3'(B)$ curves, related to the flux creep, becomes more pronounced as the temperature increases. Nevertheless the height of the modulus peak decreases and this could be explained in terms of a reduction of the non-linearity of the $E$-$J$ characteristic, driven by temperature, due to the crossover between the flux creep and the flux flow regime.

## 5. Conclusions

The frequency dependences of the third harmonic susceptibilities have been analysed as a function of the temperature and the dc magnetic field on an $MgB_2$ bulk sample. The combined analysis of the $\chi_3(T)$ and $\chi_3(B)$ curves has allowed us to investigate the contribution of the different vortex dynamics to the magnetic response of the sample. Moreover, we have performed numerical simulations of the magnetic diffusion processes in the framework of a new empirical model. Within this approach the different regimes of the flux motion are naturally recovered. By comparing the experimental curves with the simulated ones we have resolved the crossover between a dissipative regime governed by flux creep and one dominated by flux flow phenomena. We have also shown that the frequency behaviour of the $\chi_3(B)$ curves indicates the presence of a vortex glass dynamic in the flux creep regime.

## 6. Acknowledgments

We express our thanks to Dr. Tiziana Di Matteo for helpful discussions about the numerical simulations. Thanks are also due to Dr. Danilo Zola and Dr. Bernd Seeber for useful suggestions and to Dr. Giovanni Giunchi for his $MgB_2$ samples.

**Figure captions**

**Figure 1**: Differential resistivities of a pinned superconductor corresponding to Eq.(4) (line) and to Eq.(5) (open circles); in Eq.(5) also the thermally activated phenomena are taken into account. Inset (a): *E-J* curve relative to the differential resistivity in Eq.(4). Inset (b): *E-J* curve relative to the differential resistivity in Eq.(5).

**Figure 2**: Temperature dependence of $\chi'_3$ and $\chi''_3$ measured at $B_{ac}$=4 G and $f$ =1007 Hz in zero dc field.

**Figure 3**: Temperature dependence of $\chi'_3$ and $\chi''_3$ measured at $B_{ac}$=4 G and $f$ =5007 Hz in zero dc field; the $\chi_3(T)$ curves differ significantly from those of Fig. 2 as the frequency increases.

**Figure 4**: Temperature dependence of the $\chi''_3$ curves measured at $f$ =1007, 1607, 2507 Hz in (a), $f$ = 2507, 3507, 5007 Hz in (b) and $B_{ac}$=4 G, in zero dc field; the appearance of a trough is observed for $T<T_P$ ($\chi_1''$) with increasing frequency.

**Figure 5**: Temperature dependence of the $\chi''_3$ curves calculated for $f$ =1007, 1607, 2507 Hz in (a), $f$ = 2507, 3507, 5007 Hz in (b) and $B_{ac}$=4 G, in zero dc field; the behaviour of the experimental curves for $T<T_P$ ($\chi_1''$) in Fig. 4 is well reproduced.

**Figure 6**: Dc magnetic field dependence of $\chi'_3$ (a) and $\chi''_3$ (b) measured at $T$= 20 K, $f$ =1607 Hz and $B_{ac}$= 2, 4, 8, 16 G.

**Figure 7**: Dc magnetic field dependence of $\chi'_3$ (a) and $\chi''_3$ (b) measured at $T$= 20 K, $B_{ac}$= 4 G and $f$ =1007, 1607, 3507, 5007 Hz; the height of the peaks decreases in both $\chi'_3$ and $\chi''_3$, while the depth of the troughs increases in $\chi'_3$ and decreases in $\chi''_3$ as the frequency increases. Inset of Fig. 7b: Magnification of the high field region of the $\chi''_3(B)$ curve.

**Figure 8**: Dc magnetic field dependence of $\chi'_3$ (a) and $\chi''_3$ (b) calculated for $T/T_c$= 0.5, $f$ =1607 Hz and $B_{ac}$= 2, 4, 8, 16 G.

**Figure 9**: Dc magnetic field dependence of $\chi'_3$ (a) and $\chi''_3$ (b) calculated for $T/T_c$= 0.5, $B_{ac}$= 4 G and $f$ =1007, 1607, 3507, 5007 Hz; the height of the peaks increases in both $\chi'_3$ and $\chi''_3$, while the depth of the troughs decreases in $\chi'_3$ and increases in $\chi''_3$ as the frequency increases. Inset of Fig. 9b: Magnification of the high field region of the $\chi''_3(B)$ curve.

**Figure 10**: Dc magnetic field dependence of $|\chi_3|$ measured at $B_{ac}$= 4 G, $f$ =1607 Hz and $T$= 20, 25, 30 K; the height of the modulus peak decreases as the temperature increases. Inset: $\chi'_3(B)$ curves measured at $B_{ac}$= 4 G, $f$ =1607 Hz and $T$= 20, 25, 30 K.

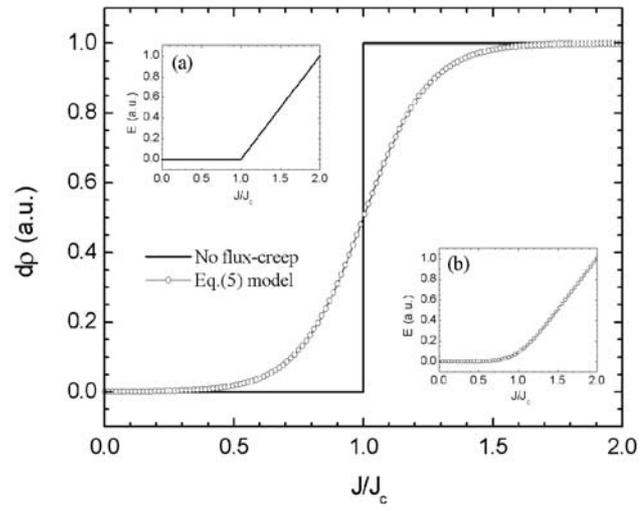

**Figure 1**

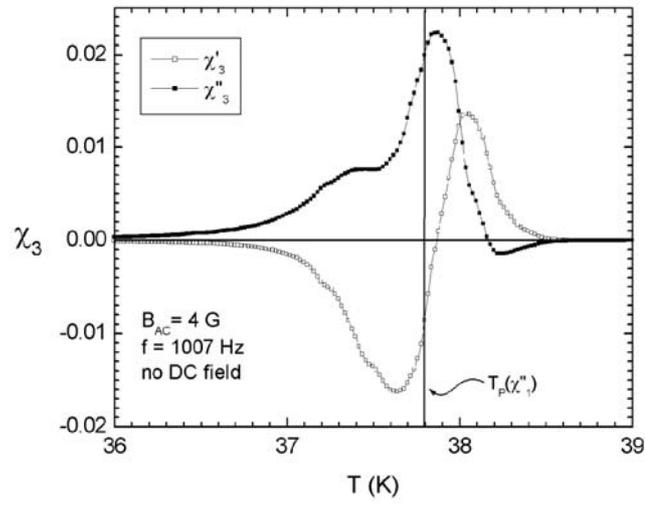

**Figure 2**

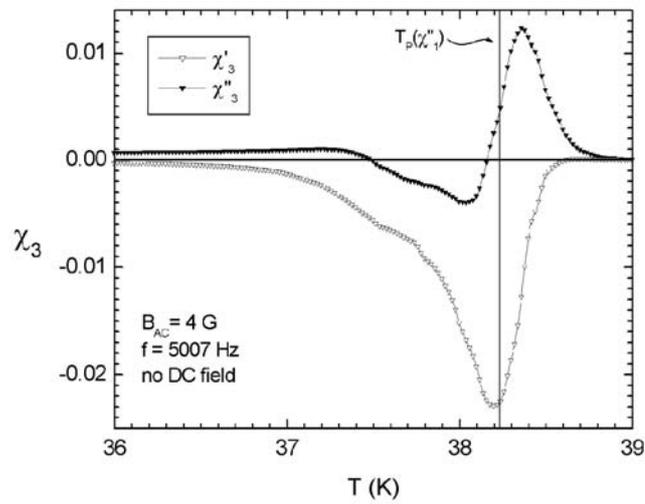

**Figure 3**

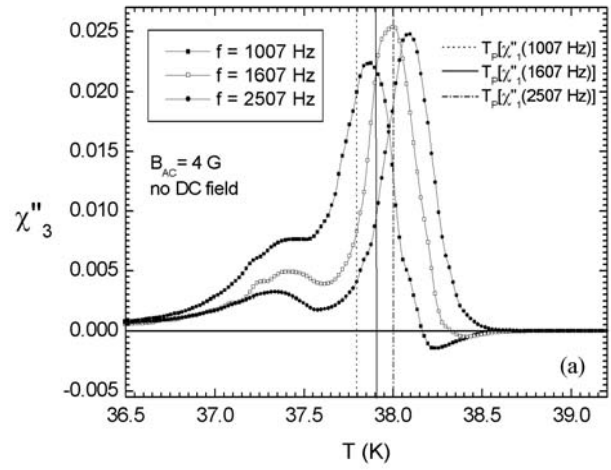

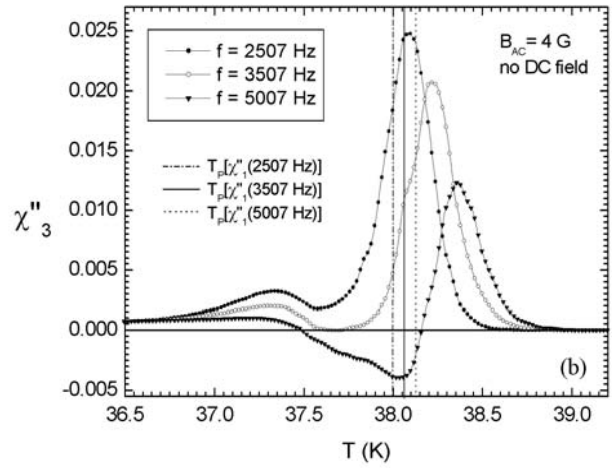

**Figure 4**

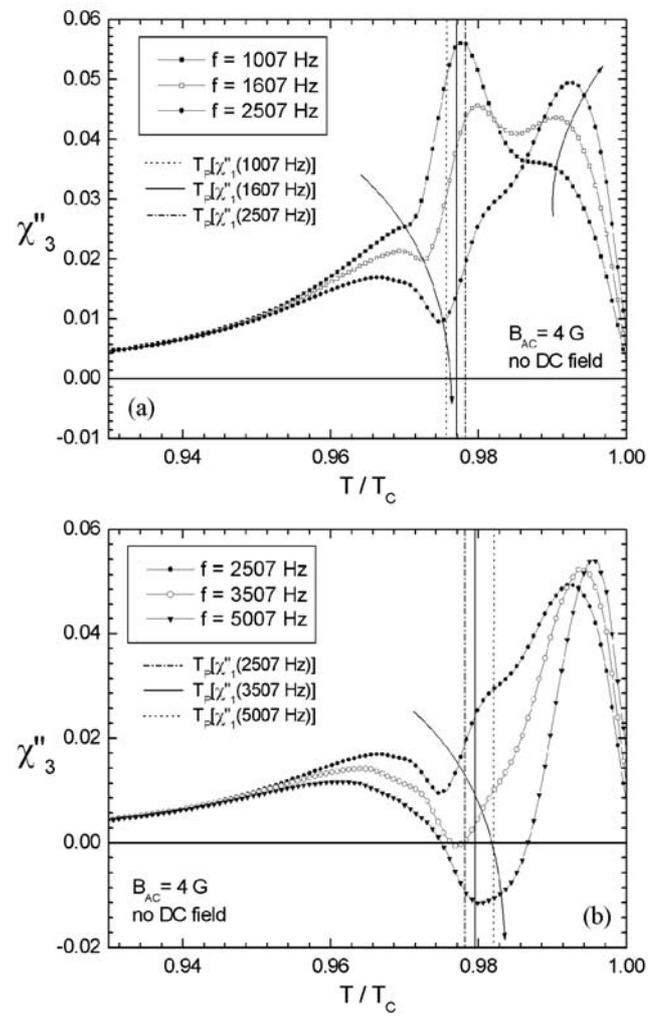

**Figure 5**

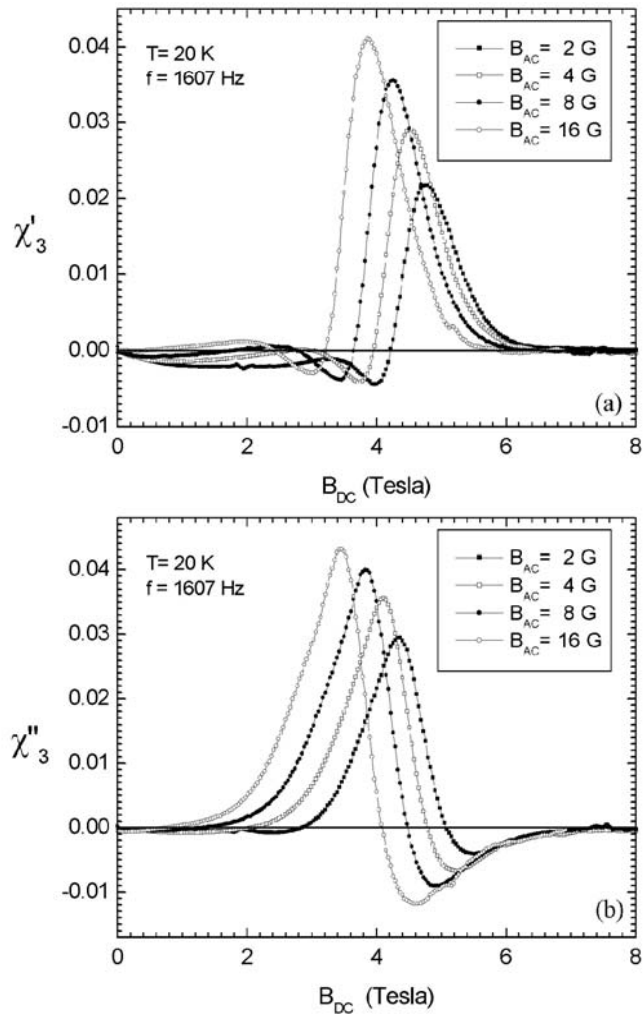

**Figure 6**

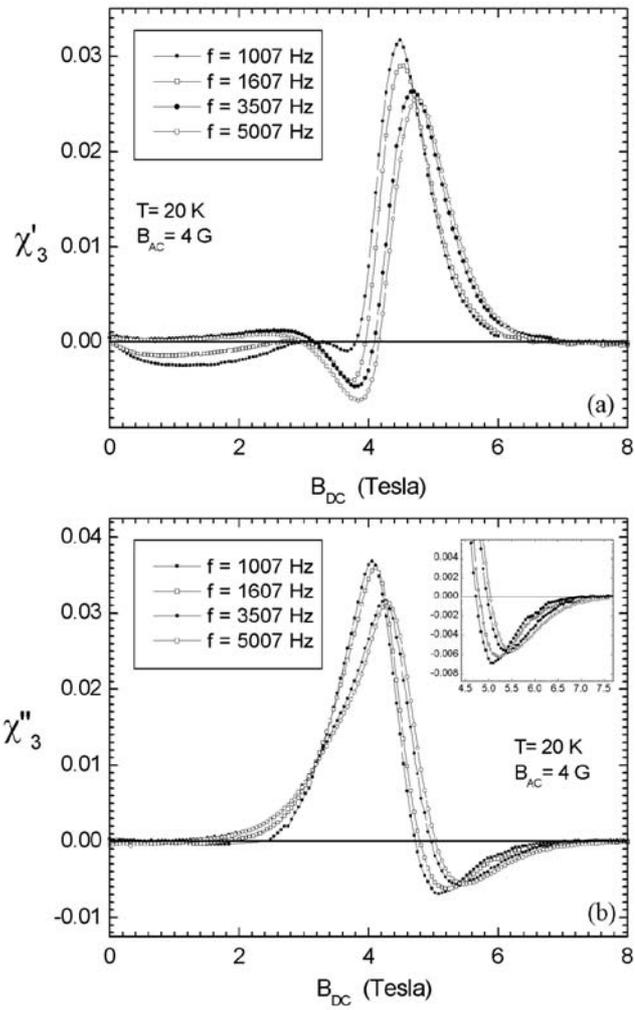

**Figure 7**

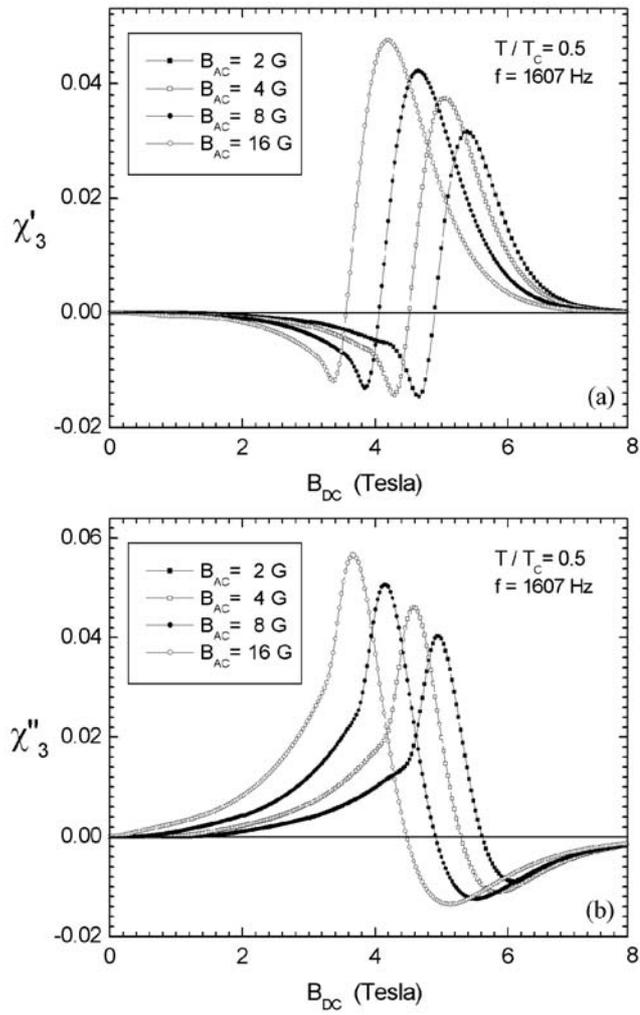

**Figure 8**

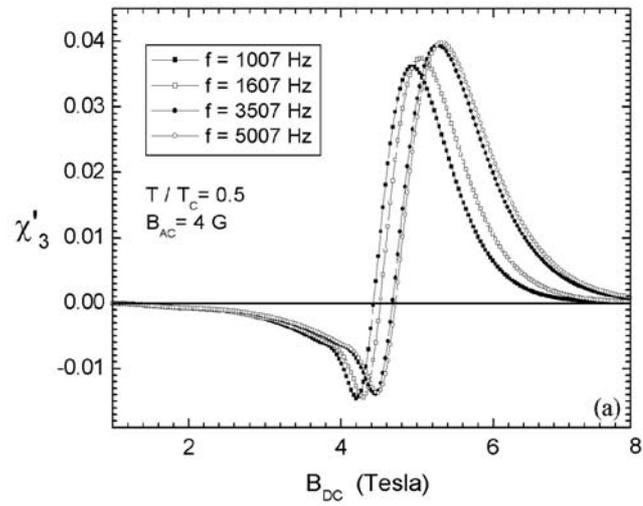

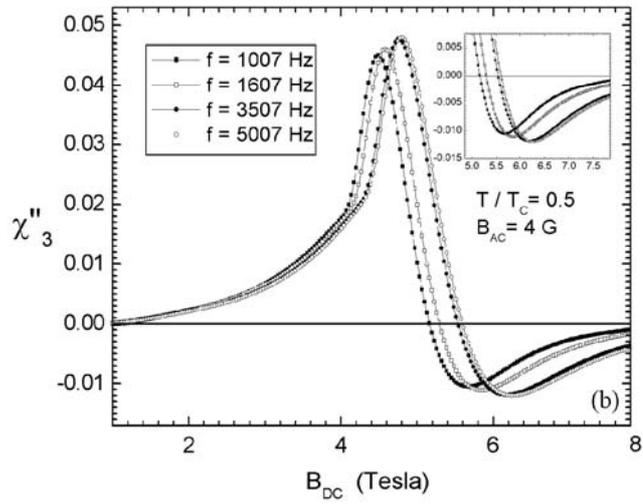

**Figure 9**

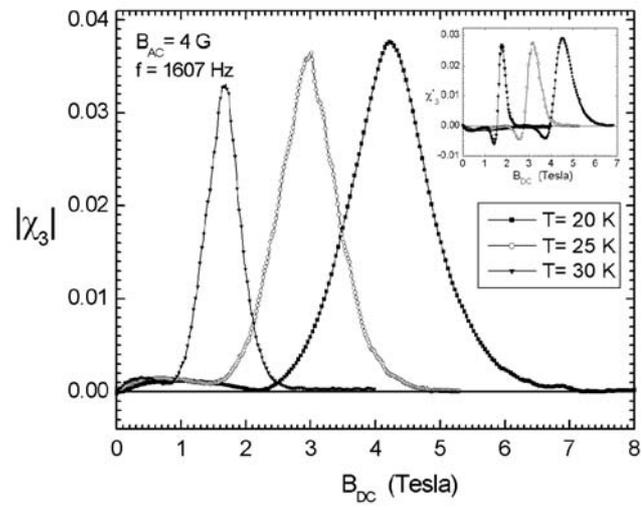

**Figure 10**